\pdfoutput=1
\documentclass[11pt]{article}

\usepackage{acl} % Use this for the final version

\usepackage{times}
\usepackage{latexsym}
\usepackage[T1]{fontenc}
\usepackage[utf8]{inputenc}
\usepackage{microtype}

\setcitestyle{numbers}

\usepackage{url}
\usepackage{booktabs}
\usepackage{graphicx}
\usepackage{amsmath}
\usepackage{tabularx}
\usepackage{ragged2e}
\usepackage{enumitem}
\usepackage{siunitx}
\usepackage{listings}
\usepackage{lmodern}        % 使用更现代的字体
\usepackage{titling}        % 用于支持副标题

\setcitestyle{authoryear}

\hypersetup{
    colorlinks=true,
    allcolors=blue,
}

\title{
    THOR: Transformer Heuristics for On-Demand Retrieval \\ % 主标题后换行
    \vspace{0.5em} % 增加一点垂直间距，让布局更好看
    \normalsize An LLM Solution Enabling Conversation with Relational Databases by eSapiens % 副标题，使用\large字号
}

\author{Isaac Shi \and Zeyuan Li \and Fan Liu \and Wenli Wang \\
        \bf{Lewei He \and Yang Yang \and Tianyu Shi} \\
  eSapiens Team \\
  \texttt{\{ishi, zeyuanli, fanliu, wenliwang, leweihe, yangyang, tianyushi\}@esapiens.ai} \\}

\begin{document}
\maketitle

\begin{abstract}
We introduce the \textbf{THOR (Transformer Heuristics for On-Demand Retrieval) Module} Designed and implemented by eSapiens, a secure, scalable engine that transforms natural-language questions into verified, read-only SQL analytics for enterprise databases. The Text-to-SQL module follows a decoupled \emph{orchestration/execution} architecture: a Supervisor Agent routes queries, Schema Retrieval dynamically injects table and column metadata, and a SQL Generation Agent emits single-statement \texttt{SELECT} queries protected by a read-only guardrail. An integrated Self-Correction \& Rating loop captures empty results, execution errors, or low-quality outputs and triggers up to five LLM-driven regeneration attempts. Finally, a Result Interpretation Agent produces concise, human-readable insights and hands raw rows to the Insight \& Intelligence engine for visualization or forecasting.

Smoke tests across finance, sales, and operations scenarios demonstrate reliable ad-hoc querying and automated periodic reporting. By embedding schema awareness, fault-tolerant execution, and compliance guardrails, the THOR Module empowers non-technical users to access live data with zero-SQL simplicity and enterprise-grade safety.More details and demos are available at: \url{https://www.esapiens.ai/}.
\end{abstract}

\section{Introduction}

\noindent
Many business questions still require on-the-fly SQL, yet manual querying is slow and error-prone. The \textbf{THOR Module} converts natural-language requests into verified database results and then hands those results to the \emph{Insight \& Intelligence} engine for downstream visualisation or forecasting through a multi-agent pipeline. The process begins with a Supervisor Agent that handles task routing, detecting numeric-analysis intent and dispatching the request to the T2S lane. To ensure the language model has the necessary context, the system performs on-demand schema injection, fetching relevant table and column names and embedding them into the prompt. The SQL generation phase uses GPT-4o to produce a single \emph{read-only} statement, which inherently blocks any potentially harmful update operations. A critical execution and self-correction loop follows; if the initial query fails, returns an empty set, or receives a poor quality score, the system attempts to fix the SQL and retries up to five times. Finally, the raw data rows are handed off to the user and streamed to the Insight \& Intelligence engine, which generates the chart, forecast, or narrative specified by the calling workflow.

\subsection*{Contributions}
Our primary contribution is an enterprise-ready Text-to-SQL system that prioritizes safety, robustness, and usability. We demonstrate its effectiveness through a series of tests on realistic business queries. Key findings include:
\begin{itemize}[leftmargin=*]
  \item \textbf{Self-healing pipeline}: A bounded retry loop salvages queries that would otherwise fail, significantly improving the success rate over baseline agents.
  \item \textbf{Semantic guardrails}: Built-in heuristics for units, dates, and fuzzy text matching prevent common “looks-right-but-wrong” answers.
  \item \textbf{Readable insights}: The system appends a short narrative to query results, making them immediately actionable for non-technical users.
  \item \textbf{Compliance by construction}: All queries are enforced as read-only, and the system correctly refuses unauthorized column requests, ensuring data governance.
\end{itemize}

The current release has passed smoke tests on finance, sales, and operations scenarios and reliably answers everyday queries while inheriting the platform’s read-only access model and audit logging.

\section{Motivation}
\label{sec:motivation}

\begin{figure*}[htbp]
    \centering
    \includegraphics[width=0.9\textwidth]{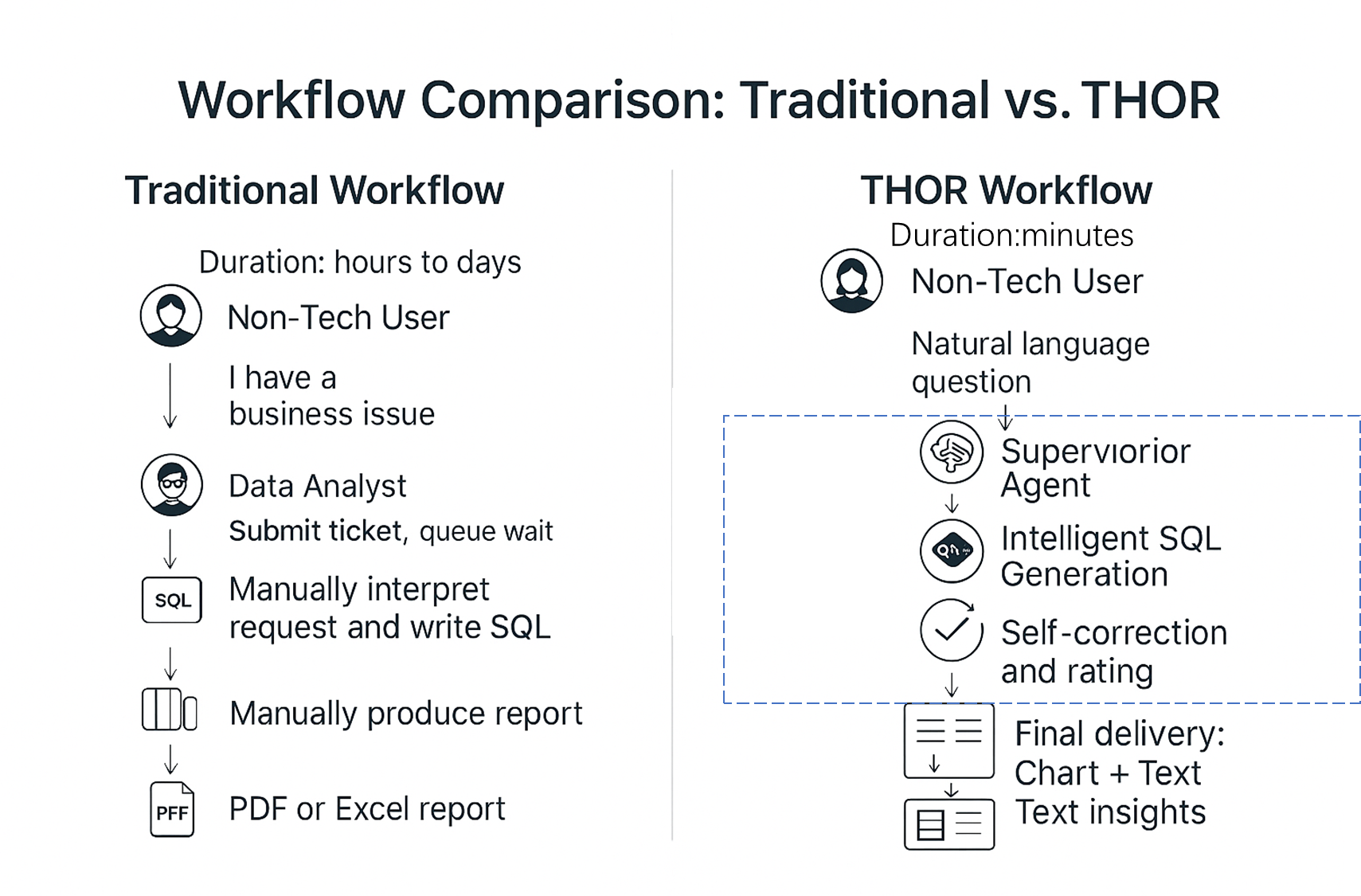}
    \caption{Comparison of a traditional data analysis workflow (left), characterized by manual handoffs and multi-day turnaround times, versus the streamlined, automated workflow of the THOR module (right), which delivers insights directly to non-technical users in minutes.}
    \label{fig:workflow_comparison}
\end{figure*}

%%% --- MODIFICATION START --- %%%
Despite the proliferation of Business Intelligence (BI) tools, enabling frontline teams to directly interrogate structured relational databases in plain English remains a significant challenge. This difficulty stems from the inherent semantic chasm between intuitive, human-language inquiries and the rigid, formal syntax of SQL. The THOR module is designed to bridge this gap by addressing three systemic obstacles that hinder the democratization of structured data access.
%%% --- MODIFICATION END --- %%%
%%% --- MODIFICATION START --- %%%
Figure~\ref{fig:workflow_comparison} starkly illustrates this contrast, comparing the slow, manual, and multi-step traditional process with the direct, automated query-to-insight path provided by THOR.

%%% --- MODIFICATION END --- %%%
\subsection{Analytics Skill Gap}
Business users can frame a question but rarely write SQL. Data teams become “report factories,” juggling clarification meetings, ticket queues, and re-work. The turnaround from question to answer stretches from hours to days, slowing operational decisions.

\subsection{Dashboard Rigidity \& the Ad-Hoc Blind Spot}
Conventional dashboards excel at fixed KPIs but break down when users need one-off joins, custom filters, or exploratory slices. The fallback is exporting CSVs for manual crunching or filing new data tickets—both slow and labour-intensive.

\subsection{Data Compliance \& Read-Only Safety}
Letting an LLM touch production databases introduces two distinct risks. First is the write risk, where a hallucinated \texttt{UPDATE} or \texttt{DELETE} statement could corrupt live data. Second is the compliance risk, where query results may expose rows or columns that should remain shielded by row- or column-level security policies. Without enforced read-only execution and fine-grained auditing, most NL-to-SQL prototypes fail compliance review. The THOR module tackles these gaps through a closed-loop system that removes the skill barrier, enables true ad-hoc queries, and embeds compliance guardrails from the ground up.

\begin{table*}[!htb]
\centering
\renewcommand{\arraystretch}{1.15}
\setlength{\tabcolsep}{6pt}
\begin{tabularx}{\textwidth}{@{}
  >{\raggedright\arraybackslash}X
  >{\raggedright\arraybackslash}X
  >{\raggedright\arraybackslash}X
@{}}
\toprule
\textbf{Prompt theme} &
\textbf{Typical pitfall for baseline agents} &
\textbf{How eSapiens handled it} \\
\midrule
Non-existent status codes &
Agent-A returned an empty table because it queried \texttt{status LIKE '\%pending\%'}, a value that does not exist. &
First attempt also failed; the introspection loop fetched the real status list, regenerated SQL, and produced correct counts. \\[4pt]

Unit conversion (metre $\rightarrow$ mile) &
All baselines echoed the raw distance column, ignoring the requested miles conversion. &
Injected \texttt{distance/1609.34} directly into SQL, returning correct rankings. \\[4pt]

Fuzzy genre match (“hip-hop”) &
Agents B \& C required an exact \texttt{'Hip Hop'} match, yielding zero rows. &
On retry, switched to \texttt{ILIKE '\%hip\%hop\%'} and succeeded. \\[4pt]

Future-dated rows in “last-3-months” query &
Baselines silently included future timestamps. &
Rating layer flagged the anomaly; SQL was regenerated with a strict date predicate. \\[4pt]

Multi-step request (top albums + buyer country) &
Agent-B mis-parsed the join intent; Agent-C declined to answer. &
Generated the two-step SQL correctly and merged the results into one answer. \\
\bottomrule
\end{tabularx}
\caption{Comparison of THOR and baseline agents on common real-world query challenges. Full SQL and results are in Appendix A.}
\label{tab:practice}
\end{table*}

\section{Related Work}
\label{sec:t2s_related}

Early neural approaches laid the groundwork for modern Text-to-SQL systems. \textsc{Seq2SQL}, for instance, introduced policy-gradient optimisation that rewards executable queries and pioneered sketch-based decoding for simple table settings \citep{zhong2017seq2sql}. Following this, SQLNet removed explicit grammar supervision by predicting column and value sets, which improved convergence on the WikiSQL dataset without relying on reinforcement learning \citep{xu2017sqlnet}.

The development of cross-domain benchmarks and structure-aware encoders marked a significant step forward. The \textsc{Spider} dataset, in particular, established a large, human-annotated benchmark with complex, cross-database queries that remains the de-facto testbed for Text-to-SQL research \citep{yu2018spider}. To tackle Spider’s schema heterogeneity, RAT\textsubscript{SQL} augmented encoder representations with relation-aware graph links to achieve state-of-the-art exact-match accuracy above 70\,\% \citep{wang-etal-2020-rat}. PICARD further improved on this by constraining autoregressive decoding, validating partial SQL against the database grammar at each timestep to largely eliminate syntax errors \citep{scholak2021picard}.

More recently, research has shifted towards large language model (LLM) baselines and advanced prompt engineering. Rajkumar et al. demonstrated that GPT-3.5 can attain competitive scores on the Spider benchmark in zero- or few-shot setups, provided that schema and example selection are carefully crafted \citep{rajkumar2022evaluatingtexttosqlcapabilitieslarge}. Gao et al. extended this line of work with \emph{DAIL-SQL}, a benchmark evaluation that combines condensed exemplars with GPT-4 to push execution accuracy beyond 85\% while controlling for prompt length \citep{gao2023dailsql}.

Recent surveys and trend analyses help to contextualize these advancements. One comprehensive survey summarises the rapid progress in LLM-enhanced Text-to-SQL generation, covering key areas such as prompt design, schema retrieval, execution feedback, and agent-style pipelines. It also outlines open challenges in cost control and compliance that remain relevant to production systems \citep{llmtsqlsurvey2024}.

\smallskip
%%% --- MODIFICATION START --- %%%
Together, these studies chart the evolution from early reinforcement-learning parsers to sophisticated, LLM-driven prompting techniques. The THOR module synthesizes these academic advancements into an enterprise-ready framework. Rather than proposing a novel SQL generation algorithm, its primary contribution lies in the holistic system architecture that integrates state-of-the-art generation with robust, fault-tolerant execution, a semantic interpretation layer for results, and strict compliance guardrails. It operationalizes the findings of the cited research while addressing practical concerns of safety, reliability, and usability often overlooked in pure benchmark-focused studies.
%%% --- MODIFICATION END --- %%%

\section{System Architecture}
\label{sec:architecture}

\begin{figure*}[htbp]
  \centering
  \includegraphics[width=\linewidth]{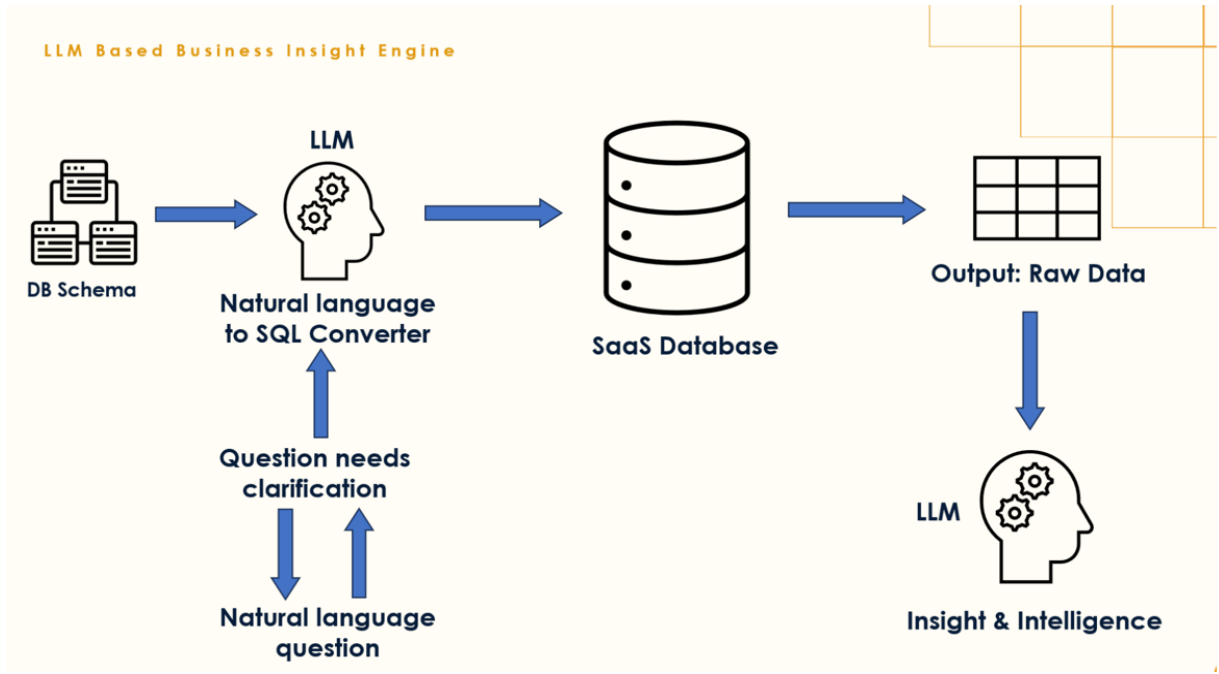}
  % --- MODIFICATION: Moved caption to the bottom ---
  \caption{Overall T2S architecture, showing the decoupled orchestration and execution layers with specialized agents for SQL generation, interpretation, and self-correction.}
  \label{fig:t2s_arch}
\end{figure*}

This module targets structured data question answering tasks and is built upon a multi-agent coordination architecture. As illustrated in Figure~\ref{fig:t2s_arch}, the system enables dynamic reasoning over structured databases by leveraging a decoupled architecture that separates orchestration from execution.

% --- MODIFICATION: Moved Figure 1 to appear near its reference text ---

It consists of a task dispatcher and several specialized functional agents:

\begin{itemize}
\item \textbf{Supervisor Agent}: Responsible for receiving user queries, interpreting task types, and routing them to appropriate functional agents.
\item \textbf{SQL Generation Agent}: Converts natural language questions into executable SQL queries based on database schema and semantic understanding.
\item \textbf{Result Interpretation Agent}: Analyzes the returned tabular data, extracts key values and trends, and generates semantic explanations in natural language.
\item \textbf{Self-correction Module}: Automatically triggered upon query failure or unsatisfactory results; re-generates SQL queries using scoring and retry mechanisms.
\end{itemize}

\subsection{Operational Workflow}

The Data Analyze Agent follows a closed-loop workflow covering question analysis, SQL generation, execution, and insight synthesis. The process consists of the following steps:

\begin{itemize}
\item \textbf{Task Analysis and Routing}: Upon receiving a natural language question, the Supervisor Agent determines whether it requires structured query processing and dispatches it to the SQL Generation Agent.
\item \textbf{SQL Construction and Execution}: The SQL Generation Agent constructs a read-only SQL statement by referencing schema structure and field semantics, then submits the query to the backend data warehouse.
\item \textbf{Self-Correction \& Rating}: If the query returns an empty result, an execution error, or a low-quality rating, the Self-Correction Module is invoked to analyze the issue and regenerate the SQL.
\item \textbf{Insight Generation}: Once a valid result is obtained, the Result Interpretation Agent analyzes the tabular output, extracts key values, and converts them into user-facing natural language insights.
\item \textbf{Answer Delivery}: The final answer is presented in a readable format, supporting either concise summarization or detailed explanatory output.
\end{itemize}

\subsection{Self-Correction \& Rating}
\label{sec:self-correction}

\begin{figure}[!htb]
  \centering
  \includegraphics[width=0.95\linewidth]{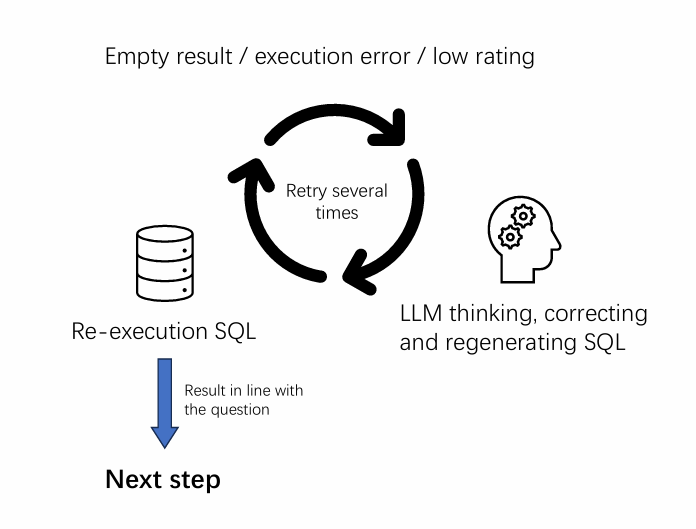}
  \caption{Self-correction \& rating process: when an empty result, execution error, or low-quality rating is detected, the system automatically regenerates and re-executes the SQL up to a fixed number of retries, guided by LLM feedback and scoring.}
  \label{fig:self-correction}
\end{figure}

To enhance SQL execution success and improve result accuracy, the self-correction workflow (Figure~\ref{fig:self-correction}) detects empty results, execution errors, or low-quality ratings. When triggered, it uses LLM feedback to re-generate the SQL query and re-executes it, with a hard limit on the number of retries to prevent infinite loops. This mechanism is critical for handling ambiguous queries and schema mismatches automatically.

\section{Experimental Evaluation}
\label{sec:evaluation}

To gauge real-world robustness, we issued seven natural-language queries to \emph{THOR} and three popular text-to-SQL agents (anonymised here as \textbf{Baseline Agents}). The prompts span both logistics and retail schemas and include common challenges like unit conversion, fuzzy text matching, and multi-step analytics. Table~\ref{tab:practice} summarises the most common failure patterns we saw in the baselines and how eSapiens resolved them. Full prompts and generated SQL are available in Appendix A.

\subsection{Case Study: Logistics Operational Oversight}

\subsubsection{Problem Scenario}
Logistics organisations contend with massive, real-time data streams from shipment-tracking systems, vendor portals, and inventory tools. Front-line teams spend countless hours compiling status reports and reacting to delays \emph{after the fact}, leading to siloed decisions and missed opportunities.

\subsubsection{Solution Implementation}
\begin{figure*}[!htb]
  \centering
  \includegraphics[width=0.9\linewidth]{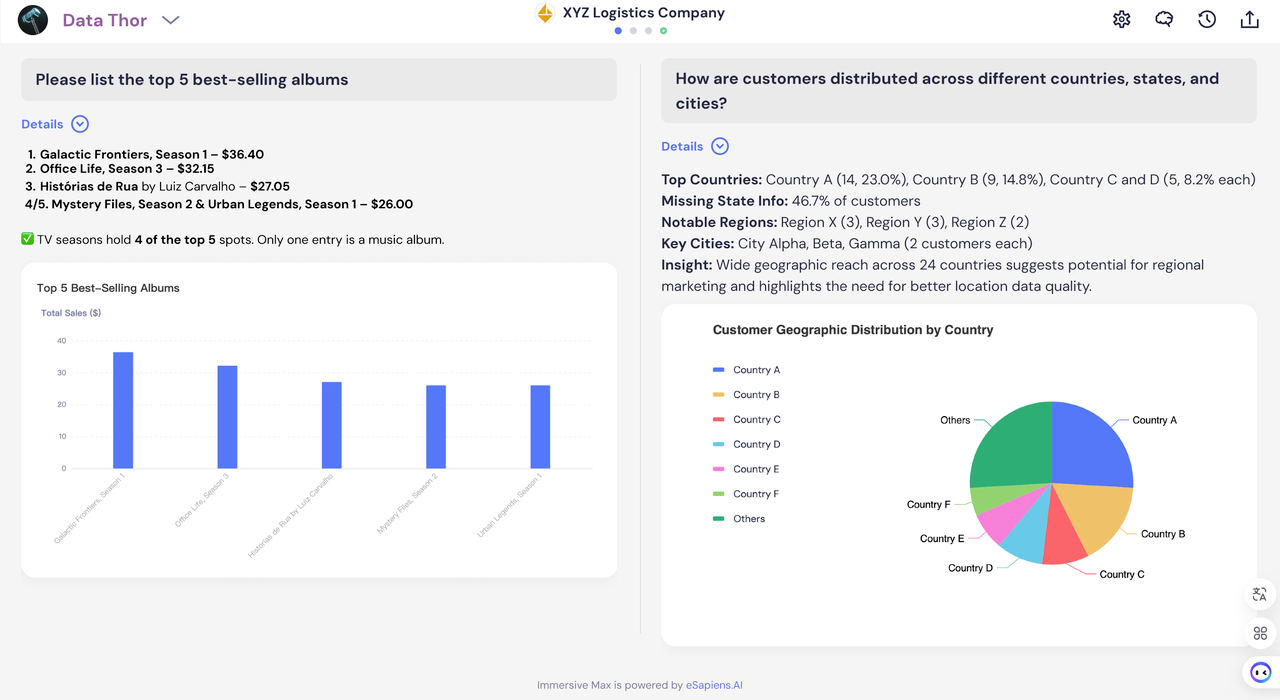}
  \caption{A natural-language query is instantly converted to read-only SQL with summarised, chart-based results, allowing for at-a-glance trend analysis.}
  \label{fig:t2s_demo}
\end{figure*}

We deployed the THOR module to provide direct, natural-language access to live logistics data. The solution centered on three key capabilities. First, it provided a natural-language interface for logistics KPIs, allowing staff to ask real-time questions such as, “What are the top 3 reasons for shipment delays last week?” The module then converts these requests into read-only SQL, as shown in Figure~\ref{fig:t2s_demo}. Second, the system enabled automated reporting by scheduling daily performance summaries and exception alerts, which significantly reduced manual effort. Third, it delivered unified data insights by connecting WMS, TMS, and ERP data into a single, searchable workspace, thereby enabling complex cross-system queries.

\subsubsection{Results and Impact}
As shown in Table~\ref{tab:key_benefits_logistics}, the T2S-powered workflow reduced routine reporting and diagnostic tasks from hours to minutes. This delivered up to \textbf{90\%+} time savings and enabled near real-time decision-making, allowing teams to proactively manage disruptions and improve throughput.

\begin{table*}[!htb]
\centering
\renewcommand{\arraystretch}{1.2}
\setlength{\tabcolsep}{6pt}
\begin{tabularx}{\textwidth}{@{}
  >{\raggedright\arraybackslash}X
  >{\centering\arraybackslash}X
  >{\centering\arraybackslash}X
  >{\centering\arraybackslash}X@{}}
\toprule
\textbf{Task} & \textbf{Manual Effort} & \textbf{With AI Agent} & \textbf{Time Saved} \\
\midrule
Delay analysis report        & $\sim$1 hour/day        & 2–3 mins/query    & $\downarrow$ 90\%         \\
Weekly performance review      & 4–6 hours/week          & Auto-generated    & $\downarrow$ 100\% (hands-off) \\
Multi-source metric comparison & Needs analyst support   & Real-time         & $\downarrow$ 80\%          \\
Escalation lead time           & Reactive (1–2 days)     & Near real-time    & $\downarrow$ 70–90\%       \\
\bottomrule
\end{tabularx}
\caption{Key benefits delivered by the logistics AI agent in the case study.}
\label{tab:key_benefits_logistics}
\end{table*}

\section{Conclusion and Future Work}

The current release of the \textbf{THOR Module} already lets non-technical users query enterprise data with plain-language prompts and receive verified, read-only results in seconds. To push the system from “mostly helpful” to “consistently correct,” our future work will focus on three practical, SQL-centric enhancements. First, we plan to implement domain schema enrichment and synonym mapping by building a lightweight dictionary that links business terms, acronyms, and metric names to their underlying tables and columns. Injecting these mappings into the prompt will help the model choose the correct field, even with informal phrasing. Second, we will introduce a graph-aware join planner that maintains an internal graph of foreign key relationships and common join paths, allowing it to propose valid join templates before query execution and reduce the risk of missing links or accidental Cartesian products. Finally, we will establish a live execution feedback loop to capture database error messages and basic result sanity checks. This information will be fed back into a nightly fine-tuning script and a growing regression suite, allowing the model to learn from real-world missteps and steadily improve its SQL accuracy.

By delivering these work streams, the THOR module will evolve into a more semantically aware, join-safe, and self-improving NL-to-SQL engine, ready for increasingly complex analytics scenarios in production.

\section{Limitations}

The THOR module's effectiveness is subject to several inherent Text-to-SQL challenges. First, its performance on queries requiring complex relational logic, such as multiple nested subqueries or advanced window functions, is not yet fully validated. Second, dynamic schema injection faces scalability issues with large enterprise data warehouses, as the LLM's finite context window cannot accommodate schemas with hundreds or thousands of tables. Finally, the system relies on the LLM's general knowledge and may misinterpret domain-specific jargon or implicit business rules without a dedicated semantic layer, potentially leading to logically valid but contextually incorrect results.

% 告诉 LaTeX 使用我们创建的 my_references.bib 文件
\bibliography{reference}

\appendix
\section*{Appendix}

\section{Natural-Language Prompts \& Full SQL Listings}
\label{app:sql}

\subsection*{Prompt 1}
\textbf{Natural-language question}\\
\emph{“Show me the pending deliveries by month for the last 18 months.”}

\textbf{SQL queries generated by Product X}
\begin{lstlisting}
SELECT
  DATE_TRUNC('month', delivery_requests.created_at) AS month,
  COUNT(delivery_requests.id)                       AS pending_deliveries
FROM payload_catalog.public.delivery_requests
WHERE delivery_requests.status ILIKE '%pending%'
  AND delivery_requests.created_at >= DATE_SUB(CURRENT_DATE(), 18 * 30)
GROUP BY month
ORDER BY month ASC;
\end{lstlisting}

\textbf{SQL queries generated by eSapiens' THOR}
\begin{lstlisting}
WITH monthly_deliveries AS (
  SELECT DATE_TRUNC('month', created_at) AS month,
         status,
         COUNT(*)                        AS delivery_count
  FROM   delivery_requests
  WHERE  created_at >= NOW() - INTERVAL '18 months'
    AND  created_at <= NOW()
    AND  status IS NOT NULL
  GROUP  BY DATE_TRUNC('month', created_at), status
)
SELECT TO_CHAR(month,'YYYY-MM') AS month,
       status,
       delivery_count
FROM   monthly_deliveries
ORDER  BY month DESC, status;
\end{lstlisting}

%-------------------------------------------------------
\subsection*{Prompt 2}
\textbf{Natural-language question}\\
\emph{“What are the different channels through which drivers learned about someone, and what is the percentage of drivers for each channel? In addition, analyze these data to provide actionable insights and recommendations for the marketing team to optimize their outreach strategy. '}

\textbf{SQL queries generated by Product X}
\begin{lstlisting}
SELECT
  users.invitation_token,
  ( COUNT(users.id) * 100.0 /
    (SELECT COUNT(*) FROM payload_catalog.public.users)
  ) AS percentage
FROM payload_catalog.public.users
GROUP BY users.invitation_token;
\end{lstlisting}

\textbf{SQL queries generated by eSapiens' THOR}
\begin{lstlisting}
WITH channel_counts AS (
  SELECT
    COALESCE(how_did_you_hear,'Not Specified') AS channel,
    COUNT(*)                                     AS driver_count,
    COUNT(*) * 100.0 /
      (SELECT COUNT(*) FROM driver_details)   AS percentage
  FROM driver_details
  GROUP BY how_did_you_hear
  ORDER BY driver_count DESC
)
SELECT channel,
       driver_count,
       ROUND(percentage,2) AS percentage
FROM   channel_counts;
\end{lstlisting}

%-------------------------------------------------------
\subsection*{Prompt 3}
\textbf{Natural-language question}\\
\emph{“I want to see the top 10 regions with higher income \textbf{per mile} by region in the past 3 months. Note: The distance field in the \texttt{delivery\_requests} table is measured in meters. '}

\textbf{SQL queries generated by Product Y}
\begin{lstlisting}
WITH recent_requests AS (
  SELECT
    dr.id,
    a.region_id,
    dr.fee_total_calculated,
    dr.distance
  FROM payload_catalog.public.delivery_requests dr
  JOIN payload_catalog.public.accounts          a
       ON dr.account_id = a.id
  WHERE dr.created_at >= DATE_SUB(CURRENT_DATE(), 90)
),
region_income AS (
  SELECT
    region_id,
    SUM(fee_total_calculated) /
    SUM(distance)               AS income_per_mile
  FROM recent_requests
  GROUP BY region_id
)
SELECT *
FROM   region_income
ORDER  BY income_per_mile DESC
LIMIT 10;
\end{lstlisting}

\textbf{SQL queries generated by eSapiens' THOR}
\begin{lstlisting}
WITH region_metrics AS (
  SELECT
    r.name                                       AS region_name,
    SUM(dr.fee_total_calculated/1000)            AS total_revenue_dollars,
    SUM(dr.distance / 1609.34)                   AS total_distance_miles,
    CASE
      WHEN SUM(dr.distance / 1609.34) = 0
      THEN 0
      ELSE SUM(dr.fee_total_calculated/1000) /
           SUM(dr.distance / 1609.34)
    END                                          AS revenue_per_mile
  FROM delivery_requests dr
  JOIN accounts          a ON dr.account_id = a.id
  JOIN regions           r ON a.region_id   = r.id
  WHERE dr.created_at >= CURRENT_DATE - INTERVAL '3 months'
    AND dr.status IN ('delivered','completed','DELIVERED','COMPLETED')
  GROUP BY r.name
  HAVING SUM(dr.distance) > 0
)
SELECT region_name,
       ROUND(total_revenue_dollars,2),
       ROUND(total_distance_miles,2),
       ROUND(revenue_per_mile,2) AS dollars_per_mile
FROM   region_metrics
ORDER  BY revenue_per_mile DESC
LIMIT 10;
\end{lstlisting}

%-------------------------------------------------------
\subsection*{Prompt 4}
\textbf{Natural-language question}\\
\emph{“Which track has the highest unit price?”}

\textbf{SQL queries generated by Product Y}
\begin{lstlisting}
SELECT name, unit_price
FROM   chinook_track
ORDER  BY unit_price DESC
LIMIT 1;
\end{lstlisting}

\textbf{SQL queries generated by eSapiens' THOR}
\begin{lstlisting}
-- identical query; eSapiens adds a narrative explanation
SELECT name, unit_price
FROM   chinook_track
ORDER  BY unit_price DESC
LIMIT 1;
\end{lstlisting}

%-------------------------------------------------------
\subsection*{Prompt 5}
\textbf{Natural-language question}\\
\emph{“How many hip-hop tracks are there in the database?” (requires fuzzy matching).}

\textbf{SQL queries generated by Product Z}
\begin{lstlisting}
SELECT COUNT(*) AS hip_hop_tracks
FROM   chinook_track
WHERE  LOWER(genre) = 'hip hop';
\end{lstlisting}

\textbf{SQL queries generated by eSapiens' THOR}
\begin{lstlisting}
SELECT COUNT(*) AS hip_hop_track_count
FROM   chinook_track
WHERE  LOWER(genre) LIKE '%hiphop%'
   OR  LOWER(genre) LIKE '%hip hop%'
   OR  LOWER(genre) LIKE '%hip-hop%';
\end{lstlisting}

%-------------------------------------------------------
\subsection*{Prompt 6}
\textbf{Natural-language question}\\
\emph{“Provide the sales data for the past three months.”
Note: There is future sales data in the dataset.}

\textbf{SQL queries generated by Product Z}
\begin{lstlisting}
SELECT DATE_TRUNC('month', invoice_date) AS month,
       SUM(total)                        AS total_sales
FROM   chinook_invoice
WHERE  invoice_date >= (CURRENT_TIMESTAMP - INTERVAL '3 months')
GROUP  BY month
ORDER  BY month DESC;
\end{lstlisting}

\textbf{SQL queries generated by Product Z}
\begin{lstlisting}
WITH __chinook_invoice AS (
  SELECT DATE_TRUNC('MONTH', invoice_date) AS month,
         SUM(total)                        AS total_sales
  FROM   test.chinook.chinook_invoice
  WHERE  invoice_date >= DATEADD(MONTH, -3, CURRENT_DATE)
  GROUP  BY DATE_TRUNC('MONTH', invoice_date)
)
SELECT *
FROM   __chinook_invoice
ORDER  BY month DESC;
\end{lstlisting}
\textbf{SQL queries generated by eSapiens' THOR}
\begin{lstlisting}
SELECT
  i.invoice_id,
  i.customer_id,
  i.invoice_date,
  i.total,
  c.first_name || ' ' || c.last_name AS customer_name,
  il.track_id,
  t.name                             AS track_name,
  il.unit_price,
  il.quantity,
  (il.unit_price * il.quantity)      AS line_total
FROM  chinook_invoice      i
JOIN  chinook_customer     c  ON i.customer_id = c.customer_id
JOIN  chinook_invoice_line il ON i.invoice_id  = il.invoice_id
LEFT JOIN chinook_track    t  ON il.track_id   = t.track_id
WHERE i.invoice_date BETWEEN '2025-01-17' AND '2025-04-17'
ORDER BY i.invoice_date DESC, i.invoice_id;
\end{lstlisting}

%-------------------------------------------------------

\end{document}